\documentclass[journal]{IEEEtran}

\usepackage{theorem}
\theorembodyfont{\rmfamily}
\newtheorem{Lem}{Lemma}[section]
\newtheorem{Def}[Lem]{Definition}
\newtheorem{The}[Lem]{Theorem}
\newtheorem{Prop}[Lem]{Proposition}

\newtheorem{Rem}[Lem]{Remark}

\newtheorem{Ax}[Lem]{Axiom}
\newcommand{\qed}{\hbox{\rule{6pt}{6pt}}}

\begin{document}

\title{On uniqueness theorems for Tsallis entropy and Tsallis relative entropy}

\author{Shigeru Furuichi,~\IEEEmembership{Member,~IEEE,}
\thanks{Manuscript received November 9 2004; revised March 21, 2005.
        This work was supported by the Japanese Ministry of Education, Science, Sports and Culture, Grant-in-Aid for 
Encouragement of Young scientists (B), 17740068.}
\thanks{S.Furuichi is with the Department of Electronics and Computer Science,
Tokyo University of Science, Yamaguchi, Sanyo-Onoda City, 756-0884, Japan. Email: furuichi@ed.yama.tus.ac.jp.}}


\maketitle
\begin{abstract}
The uniqueness theorem for Tsallis entropy was presented in {\it H.Suyari, IEEE Trans. Inform. Theory, Vol.50, pp.1783-1787 (2004)}
by introducing the generalized Shannon-Khinchin's axiom.
In the present paper, this result is generalized and simplified as follows:
{\it Generalization}: The uniqueness theorem for Tsallis relative entropy is shown by means of the generalized Hobson's axiom.
{\it Simplification}: The uniqueness theorem for Tsallis entropy is shown by means of the generalized Faddeev's axiom.
\end{abstract}
\vspace{3mm}
\vspace{3mm}



\section{Introduction}  \label{sec1}
Shannon entropy is uniquely determined by the Shannon-Khinchin's axiom \cite{Kin1}, 
which is referred to as the uniqueness theorem for Shannon entropy.
The Shannon-Khinchin's axiom was improved by A. D. Faddeev \cite{Fad} in the sense that the conditions of his axiom are simpler than those of the Shannon-Kinchin's axiom. 
(For details, see \cite{Kin2,Tve,Fei,Kak}.) 
As a generalization of the axiomatic characterization of relative entropy, the uniqueness theorem for relative entropy was proven by A. Hobson \cite{Hob}.
Moreover, recently, the nonextensive entropies, including Tsallis entropy, were characterized by H. Suyari in terms of the generalized Shannon-Khinchin's axiom \cite{Suy}. 
The uniqueness theorem obtained by a generalization of the Shannon-Khinchin's axiom for the structural $a$-entropy, 
which is one of the nonextensive entropies, was the first appearance of such generalized results \cite{HC}.

The present paper proves the uniqueness theorem for Tsallis relative entropy by combining the axioms of Hobson and Suyari. 
This uniqueness theorem is a simultaneous generalization of their results with respect to the following two points.
The Hobson's theorem is generalized as a parametric extension, and the Suyari's theorem is generalized in the sense that the relative entropy function is an extension of the entropy function.  
 In addition, the uniqueness theorem for Tsallis entropy
is proven by the simplification of the generalized Shannon-Khinchin's axiom in the sense that the Faddeev's axiom is simpler than Shannon-Khinchin's axiom. 

This paper is organized as follows. In Section \ref{sec2}, a brief review of Tsallis entropy and Tsallis relative entropy is presented.
In Section \ref{sec3}, the generalized Shannon-Khinchin's axiom introduced by H. Suyari and the uniqueness theorem for Tsallis entropy are reviewed. 
In Section \ref{sec4}, the uniqueness theorem for Tsallis relative entropy by means of the generalized Hobson's axiom is shown, and the function $\phi(q)$, which will be described in Theorem \ref{the1}, is characterized. 
In Section \ref{sec5}, the uniqueness theorem for Tsallis entropy is shown by means of the generalized Faddeev's axiom. 
Finally, a theorem on the relation among the generalized Shannon-Khinchin's axiom, the generalized Faddeev's axiom, and Tsallis entropy is presented. 


\section{Tsallis entropy and Tsallis relative entropy}  \label{sec2}
\subsection{Tsallis entropy}
Several extensions of entropy have been formulated and investigated \cite{R61,R62,Dar,AD,Tsa,AO}.  
Prevalent among these extensions is the R\'enyi entropy \cite{R61,R62,AD}:
\begin{equation}
R_q \left( X \right) = \frac{1}{{1 - q}}\log \sum\limits_{i=1}^n {p_i^q } ,\quad (q\neq 1),
\end{equation}
which has the additivity property 
\begin{equation} \label{add_renyi}
R_q(X\times Y) = R_q(X) + R_q(Y),
\end{equation}
for two independent random variables $X$ and $Y$.
Tsallis entropy introduced in \cite{Tsa}, 
the definition of which is presented herein, 
the structural $a$-entropy \cite{HC}, and the entropy of type $\beta$ introduced in \cite{Dar} are referred to as nonadditive (nonextensive) entropies
because these entropies do not have an additivity for two independent random variables, whereas the R\'enyi entropy introduced in \cite{R61} 
and Shannon entropy are referred to as additive (extensive) entropies, due to Eq.(\ref{add_renyi}). The present study examines the nonextensive entropies, 
including Tsallis entropy as a typical example. 
Tsallis entropy was defined by C. Tsallis in the field of the statistical physics for the purpose of analyzing multifractal systems \cite{Tsa}.

\begin{Def} {\bf (Tsallis \cite{Tsa})}
For any nonnegative real number $q$ and the probability distribution $p_i \equiv p(X=i), (i=1,\cdots ,n)$ of a given random variable $X$,
Tsallis entropy is defined as follows:
\begin{equation}
S_q(X) = -\sum_{i=1}^n p_i^q \ln_q p_i,
\end{equation}
with parameter $q$ as an extension of Shannon entropy, where the $q$-logarithm is defined as $\ln_q(x) \equiv \frac{x^{1-q}-1}{1-q}$ for any nonnegative real numbers $q$ and $x$,
and the convention $0 \ln_q (\cdot) \equiv 0$ is set.
\end{Def}
The Tsallis entropy $S_q(X)$ converges to Shannon entropy $-\sum_{i=1}^n p_i \log p_i$ as $q \to 1$, 
because the $q$-logarithm uniformly converges to a natural logarithm as $q \to 1$. 
Tsallis entropy plays an essential role in nonextensive statistics, which is often called Tsallis statistics, so that
several important findings have been published \cite{AO}.
In addition, since the $q$-logarithm function $\ln_q(x)$ has the pseudoadditivity property for $q \neq 1$:
\begin{equation} 
\ln_q(xy) = \ln_q(x)+\ln_q(y) +(1-q) \ln_q(x)\ln_q(y), 
\end{equation}
Tsallis entropy has the pseudoadditivity property for $q \neq 1$:
\begin{equation}  \label{pseudoadditivity_first}
S_q(X\times Y) = S_q(X)+S_q(Y) +(1-q)S_q(X)S_q(Y).
\end{equation}
\begin{Rem}
Although a simple transformation exists between the Tsallis entropy and the  R\'enyi entropy, such that
\begin{equation} \label{trans}
R_q \left( X \right) = \frac{1}{{1 - q}}\log \left\{ {1 + \left( {1 - q} \right)S_q \left( X \right)} \right\},
\end{equation} 
their structures differ with respect to their additivities (cf. Eq.(\ref{add_renyi}) and Eq.(\ref{pseudoadditivity_first})).
\end{Rem}


\subsection{Tsallis relative entropy}
A relative entropy based on Tsallis entropy which is a typical nonextensive entropy, was formulated and discussed in \cite{Shi,Tsa2,RA} from a physical point of view.
The fundamental properties of Tsallis relative entropy in both classical and quantum systems were investigated in \cite{FYK} from a mathematical point of view.
\begin{Def}
Here, probability distributions $a_j$ and $b_j$ are assumed to satisfy $a_j \geq 0, b_j \geq 0$ and $\sum_{j=1}^n a_j = \sum_{j=1}^n b_j =1$.
Tsallis relative entropy between $A=\left\{a_j\right\}$ and $B=\left\{b_j\right\}$, for any $q \geq 0$, is then defined as
\begin{equation}   \label{Tsallisrelative}
D_q(A\vert B) \equiv - \sum_{j=1}^n a_j \ln_q \frac{b_j}{a_j}.
\end{equation}
\end{Def}
Note that $\lim_{q\to 1} D_q (A\vert B) = D_1(A\vert B) \equiv \sum_{j=1}^n a_j \log \frac{a_j}{b_j}$, which is known as relative entropy (often referred to as Kullback-Leibler information, divergence or cross entropy).
For Tsallis relative entropy, several fundamental properties, which are listed below, hold as parametric extensions of relative entropy (for example, \cite{FYK}). 

\begin{Prop}\label{prop_1}
\begin{itemize}
\item[(1)] (Continuity) $D_q(A\vert B)$ is a continuous function for $a_j$ and $b_j$.
\item[(2)] (Nonnegativity) $D_q(A\vert B) \geq 0$.
\item[(3)] (Symmetry) 
\begin{eqnarray*} && D_q \left( {a_{\pi \left( 1 \right)} , \cdots ,a_{\pi \left( n \right)} \left| {b_{\pi \left( 1 \right)} , \cdots ,b_{\pi \left( n \right)} } \right.} \right)\\
&& \hspace*{8mm} = D_q \left( {a_1 , \cdots ,a_n \left| {b_1 , \cdots ,b_n } \right.} \right).
\end{eqnarray*}
\item[(4)] (Possibility of extension) 
\begin{eqnarray*} && D_q \left( {a_1 , \cdots ,a_n ,0\left| {b_1 , \cdots ,b_n ,0} \right.} \right) \\
&&  \hspace*{8mm} = D_q \left( {a_1 , \cdots ,a_n \left| {b_1 , \cdots ,b_n } \right.} \right).
\end{eqnarray*}
\item[(5)] (Pseudoadditivity) 
\begin{eqnarray*} 
&&  \hspace*{-8mm} D_q \left( {A^{\left( 1 \right)}  \times A^{\left( 2 \right)} \left| {B^{\left( 1 \right)}  \times B^{\left( 2 \right)} } \right.} \right) \\
&&= D_q \left( {A^{\left( 1 \right)} \left| {B^{\left( 1 \right)} } \right.} \right) + D_q \left( {A^{\left( 2 \right)} \left| {B^{\left( 2 \right)} } \right.} \right) \\
&&+ \left( {q - 1} \right)D_q \left( {A^{\left( 1 \right)} \left| {B^{\left( 1 \right)} } \right.} \right)D_q \left( {A^{\left( 2 \right)} \left| {B^{\left( 2 \right)} } \right.} \right),
\end{eqnarray*}
where $$A^{\left( 1 \right)}  \times A^{\left( 2 \right)}  = \left\{ {a_j^{\left( 1 \right)} a_j^{\left( 2 \right)} \left|
 {a_j^{\left( i \right)}  \in A^{\left( i \right)}, i=1,2  } \right.} \right\},$$
and
$$B^{\left( 1 \right)}  \times B^{\left( 2 \right)}  = \left\{ {b_j^{\left( 1 \right)} b_j^{\left( 2 \right)} \left| 
{b_j^{\left( i \right)}  \in B^{\left( i \right)}, i=1,2} \right.} \right\}.$$
\item[(6)] (Joint convexity) For $0 \leq \lambda \leq 1$, any $q \geq 0$ and the probability distributions $A^{(i)}=\left\{a_j^{(i)}\right\}$,$B^{(i)}=\left\{b_j^{(i)}\right\}$, $(i=1,2)$,  we have 
\begin{eqnarray*} 
&& \hspace*{-8mm}  D_q \left( {\lambda A^{\left( 1 \right)}  + \left( {1 - \lambda } \right)A^{\left( 2 \right)} |\lambda B^{\left( 1 \right)}  + \left( {1 - \lambda } \right)B^{\left( 2 \right)} } \right) \\
&& \le \lambda D_q \left( {A^{\left( 1 \right)} |B^{\left( 1 \right)} } \right) + \left( {1 - \lambda } \right)D_q \left( {A^{\left( 2 \right)} |B^{\left( 2 \right)} } \right).
\end{eqnarray*}
\item[(7)] (Additivity) 

\begin{eqnarray*}
&& D_q \left( a_1 , \cdots ,a_{i - 1} ,a_{i_1 } ,a_{i_2 } ,a_{i + 1} , \cdots ,a_n | \right. \\ 
&&  \hspace*{8mm} \left. b_1 , \cdots ,b_{i - 1} ,b_{i_1 } ,b_{i_2 } ,b_{i + 1} , \cdots ,b_n   \right) \\ 
&&  = D_q \left( {a_1 , \cdots ,a_n \left| {b_1 , \cdots ,b_n } \right.} \right) \\
&&  \hspace*{8mm} +  a_i^q b_i^{1 - q} D_q \left( {\frac{{a_{i_1 } }}{{a_i }},\frac{{a_{i_2 } }}{{a_i }}\left| 
{\frac{{b_{i_1 } }}{{b_i }},\frac{{b_{i_2 } }}{{b_i }}} \right.} \right), \\ 
 \end{eqnarray*}
where $a_i  = a_{i_1 }  + a_{i_2 } ,b_i  = b_{i_1 }  + b_{i_2 }.$
\item[(8)] (Monotonicity)  For a transition probability matrix $W$, 
we have
\[
D_q \left( {WA\left| {WB} \right.} \right) \le D_q \left( {A\left| B \right.} \right).
\]
\end{itemize}
\end{Prop}
Conversely, in the present paper, Tsallis relative entropy is axiomatically characterized by the generalized Hobson's axiom.

\section{Review of the generalized Shannon-Khinchin's axioms and the uniqueness theorem for Tsallis entropy}  \label{sec3}
The generalized Shannon-Khinchin's axiom introduced by H. Suyari is reviewed in the following.
The function $S_q(x_1,\cdots ,x_n)$ is assumed to be defined for the $n$-tuple $(x_1, \cdots ,x_n)$ belonging to 
$$\Delta _n \equiv \left\{ (p_1,\cdots ,p_n) \left| \sum_{i=1}^n p_i =1, p_i \geq 0 \,\,(i=1,\cdots,n)\right.\right\}$$ 
and to take a value in $\hbox{R}^+ \equiv [0,\infty )$.

\begin{Ax} {\bf (Generalized Shannon-Khinchin's axiom)}\label{gen_SK}
\begin{itemize}
\item[(G1)] {\it Continuity}: The function $$S_q : \Delta_n \to \hbox{R}^+$$ is continuous.
\item[(G2)] {\it Maximality}:
$$
S_q\left(\frac{1}{n},\cdots ,\frac{1}{n}\right)  = \max \left\{S_q(x_1,\cdots,x_n) : x_i \in \Delta_n \right\} > 0.
$$
\item[(G3)] {\it Generalized Shannon additivity}: For $x_{ij} \geq 0$, $x_i = \sum_{j=1}^{m_i} x_{ij},\,\, (i=1,\cdots ,n ; j=1,\cdots ,m_i)$, 
\begin{eqnarray*}
&& \hspace*{-8mm} S_q(x_{11},\cdots ,x_{nm_n}) = S_q(x_1,\cdots ,x_n) \\
&&  + \sum_{i=1}^n x_i^q S_q\left(\frac{x_{i1}}{x_i},\cdots ,\frac{x_{im_i}}{x_i}\right).
\end{eqnarray*}
\item[(G4)] {\it Expandability}: $$S_q(x_1,\cdots ,x_n,0) = S_q(x_1,\cdots ,x_n).$$
\end{itemize}
\end{Ax}
Note that condition (G4) is altered slightly from [GSK4] of the original axiom \cite{Suy}.
In condition (G2), the strict positivity of the maximum value of Tsallis entropy $S_q(x_1,\cdots,x_n)$ is also imposed, in addition to [GSK2] of the original axiom \cite{Suy}.
This adoption excludes a trivial situation that Tsallis entropy is constant zero.
Then, the following theorem was shown by H. Suyari \cite{Suy}.
\begin{The} {\bf (Suyari \cite{Suy})}
Conditions (G1) to (G4) determine the function $S_q$ such that
\begin{equation}
S_q(x_1,\cdots,x_n) = \frac{1-\sum_{i=1}^n x_i^q}{\phi(q)},
\end{equation}
where $\phi(q)$ is characterized by the following conditions:
\begin{itemize}
\item[(i)] $\phi(q)$ is continuous and $\phi(q) (q-1) >0$ for $q \neq 1$.
\item[(ii)] $\lim_{q \to 1} \phi(q) = 0 $ and $\phi(q) \neq 0$ for $q \neq 1$.
\item[(iii)] There exists $(a,b) \subset \hbox{R}^+ $ such that $a < 1 <b$ and $\phi(q)$
is differentiable on $(a,1)$ and $(1,b)$. 
\item[(iv)] There exists a positive constant number $k$ such that $\lim_{q \to 1}\frac{d\phi(q)}{dq} = \frac{1}{k}$.
\end{itemize}
\end{The}

\section{Axiomatic characterization of Tsallis relative entropy by the generalized Hobson's axiom}  \label{sec4}
\subsection{Uniqueness theorem for Tsallis relative entropy}
The uniqueness theorem for relative entropy was shown by A. Hobson as follows \cite{Hob}:
\begin{The}{\bf (Hobson \cite{Hob})}\label{axiom_Hob}
The function $D_1(A\vert B)$ is assumed to be defined for any two probability distributions $A=\left\{a_j\right\}$ and $B=\left\{b_j\right\}$ for $j=1,\cdots ,n$.
If $D_1(A\vert B)$ satisfies the following conditions, then it is given by the form $k \sum_{j=1}^na_j\log\frac{a_j}{b_j}$ with a positive constant $k$.
\begin{itemize}
\item[(H1)] {\it Continuity}: $D_1(A\vert B)$ is a continuous function of $2n$ variables.
\item[(H2)] {\it Symmetry}: 
\begin{eqnarray*}
&&\hspace*{-18mm}  D_1\left(   a_1,\cdots ,a_j,\cdots ,a_k,\cdots ,a_n \vert   b_1,\cdots ,b_j,\cdots ,b_k,\cdots ,b_n   \right) \\
&&\hspace*{-18mm}  =D_1\left(   a_1,\cdots ,a_k,\cdots ,a_j,\cdots ,a_n  \vert   b_1,\cdots ,b_k,\cdots ,b_j,\cdots ,b_n   \right). 
\end{eqnarray*}
\item[(H3)] {\it Grouping axiom}:
\begin{eqnarray*}
 && \hspace*{-8mm} D_1 \left( a_{1,1} , \cdots ,a_{1,m} ,a_{2,1} , \cdots ,a_{2,m} \right. \\ 
&& \left. \hspace*{1mm}   | b_{1,1} , \cdots ,b_{1,m} ,b_{2,1} , \cdots ,b_{2,m}  \right) \\ 
&& = D_1 \left( {c_1 ,c_2 \left| {d_1 ,d_2 } \right.} \right) \\ 
 && + c_1 D_1 \left( {\frac{{a_{1,1} }}{{c_1 }}, \cdots ,\frac{{a_{1,m} }}{{c_1 }}\left| {\frac{{b_{1,1} }}{{d_1 }}, \cdots ,\frac{{b_{1,m} }}{{d_1 }}} \right.} \right)\\
&& + c_2  D_1 \left( {\frac{{a_{2,1} }}{{c_2 }}, \cdots ,\frac{{a_{2,m} }}{{c_2 }}\left| {\frac{{b_{2,1} }}{{d_2 }}, \cdots ,\frac{{b_{2,m} }}{{d_2 }}} \right.} \right)  \label{Hob_H3}
 \end{eqnarray*}
where $c_i = \sum_{j=1}^m a_{i,j}$ and $d_i = \sum_{j=1}^m b_{i,j}$.
\item[(H4)]$D_1(A\vert B) =0$ if $a_j =b_j$ for all $j$.
\item[(H5)]  $D_1(\frac{1}{n},\cdots ,\frac{1}{n},0,\cdots ,0\vert \frac{1}{n_0},\cdots ,\frac{1}{n_0})$ is an increasing function of $n_0$ and a decreasing function of $n$,
for any integers $n$ and $n_0$ such that $n_0 \geq n$.
\end{itemize}
\end{The}

The function $D_q$ is defined for the probability distributions $A=\left\{a_j\right\}$ and $B=\left\{b_j\right\}$ on a finite probability space with one parameter $q \geq 0$. 
Tsallis relative entropy is characterized by means of the following triplet of the generalized conditions (R1), (R2) and (R3).  
\begin{Ax} {\bf (Generalized Hobson's axiom)} \label{g_hobson_axiom}

\begin{itemize}
\item[(R1)] {\it Continuity}: $D_q(a_1,\cdots ,a_n \vert b_1, \cdots ,b_n)$ is a continuous function of $2n$ variables.
\item[(R2)] {\it Symmetry}: 
\begin{eqnarray*}
&&\hspace*{-18mm}  D_q\left(   a_1,\cdots ,a_j,\cdots ,a_k,\cdots ,a_n \vert   b_1,\cdots ,b_j,\cdots ,b_k,\cdots ,b_n   \right) \\
&&\hspace*{-18mm}  =D_q\left(   a_1,\cdots ,a_k,\cdots ,a_j,\cdots ,a_n  \vert   b_1,\cdots ,b_k,\cdots ,b_j,\cdots ,b_n   \right). 
\end{eqnarray*}
\item[(R3)] {\it Generalized additivity}:
\begin{eqnarray}
&&\hspace*{-18mm}    D_q \left(   a_{1,1},\cdots ,a_{1,m},\cdots ,a_{n,1},\cdots ,a_{n,m} \right. \nonumber \\ 
&& \hspace*{-18mm}   \left. \hspace*{8mm}   \vert b_{1,1},  \cdots ,b_{1,m},\cdots ,b_{n,1},\cdots ,b_{n,m}   \right)  \nonumber \\
&& \hspace*{-18mm}  =  D_q(c_1,\cdots ,c_n \vert d_1 \cdots ,d_n)   \nonumber \\
&&  \hspace*{-23mm}     + \sum_{i=1}^n c_i^q d_i^{1-q} D_q\left( \frac{a_{i,1}}{c_i},\dots ,\frac{a_{i,m}}{c_i} \vert \frac{b_{i,1}}{d_i},\dots ,\frac{b_{i,m}}{d_i} \right),\label{Tsa_H3}
\end{eqnarray} 
where $c_i = \sum_{j=1}^m a_{i,j}$ and $d_i = \sum_{j=1}^m b_{i,j}$.
\end{itemize}
\end{Ax}

Then, we have the following theorem:

\begin{The}\label{the1} 
If conditions (R1), (R2) and (R3) hold, then $D_q(A\vert B)$ is given in the following form:
\begin{equation}
D_q(A\vert B) =  \frac{1- \sum_{i=1}^n a_i^q b_i^{1-q}}{\phi (q)}
\end{equation}
with a certain function $\phi(q)$. 
\end{The}

{\it Proof:}
We prove the theorem by using conditions (R1), (R2) and (R3). 
First, we define
\begin{equation}
f_q(s,t) \equiv D_q \left(\frac{1}{s}, \cdots , \frac{1}{s},0,\cdots ,0 \vert \frac{1}{t}, \cdots ,\frac{1}{t} \right)
\end{equation}
for any natural numbers $s$ and $t$ such that $t \geq s$.
From condition (R2), we have
\begin{eqnarray*}
&& D_q \left( \frac{1}{su}, \cdots ,\frac{1}{su},0, \cdots ,0, \cdots  \cdots ,\frac{1}{su}, \cdots ,\frac{1}{su}, \right. \\
&& \hspace*{12mm}  \left. 0, \cdots ,0 | \frac{1}{tv}, \cdots ,\frac{1}{tv}  \right) \\ 
&&  = D_q \left( \frac{1}{su}, \cdots ,\frac{1}{su}, \cdots  \cdots ,\frac{1}{su}, \cdots ,\frac{1}{su},\right. \\
&& \hspace*{12mm}  \left. 0, \cdots ,0, \cdots  \cdots ,0, \cdots ,0 | \frac{1}{tv}, \cdots ,\frac{1}{tv} \right). 
 \end{eqnarray*}
Also, from condition (R3), we have
\begin{eqnarray*}
&& D_q \left( \frac{1}{{su}}, \cdots ,\frac{1}{{su}},0, \cdots ,0, \cdots  \cdots ,\frac{1}{{su}}, \cdots ,\frac{1}{{su}}, \right.\\
&& \left. \hspace*{12mm}    0, \cdots ,0 | \frac{1}{{tv}}, \cdots ,\frac{1}{{tv}} \right) \\ 
&&  = D_q \left( {\frac{1}{u}, \cdots ,\frac{1}{u},0, \cdots ,0\left| {\frac{1}{v}, \cdots ,\frac{1}{v}} \right.} \right) \\
&& \hspace*{-8mm}  + u\left( {\frac{1}{u}} \right)^q \left( {\frac{1}{v}} \right)^{1 - q} D_q \left( {\frac{1}{s}, \cdots ,\frac{1}{s},0, \cdots ,0\left| {\frac{1}{t}, \cdots ,\frac{1}{t}} \right.} \right). 
\end{eqnarray*}
From the above two equations, we have
\begin{eqnarray*}
&&  \hspace*{-8mm} D_q \left( \frac{1}{su}, \cdots ,\frac{1}{su}, \cdots  \cdots ,\frac{1}{su}, \cdots ,\frac{1}{su},0, \cdots ,\right. \\
&& \left. \hspace*{8mm} 0, \cdots  \cdots ,   0, \cdots ,0 | \frac{1}{{tv}}, \cdots ,\frac{1}{{tv}} \right) \\ 
&&   \hspace*{-8mm} = D_q \left( \frac{1}{u}, \cdots ,\frac{1}{u},0, \cdots ,0 | \frac{1}{v}, \cdots ,\frac{1}{v} \right) \\
&& \hspace*{-8mm}   + u\left( \frac{1}{u} \right)^q \left( \frac{1}{v} \right)^{1 - q} D_q \left( \frac{1}{s}, \cdots ,\frac{1}{s},0, \cdots ,0| \frac{1}{t}, \cdots ,\frac{1}{t} \right). 
\end{eqnarray*}
Thus, we have
\[
f_q \left( {su,tv} \right) = f_q \left( {u,v} \right) + \left( {\frac{u}{v}} \right)^{1 - q} f_q \left( {s,t} \right).
\]
Using a method similar to that used for Lemma 5 in \cite{HC}, we find that there exists a certain function $\phi (q)$, which depends on the parameter $q$ only, such that 
\[
f_q \left( {s,t} \right) = \frac{{1 - \left( {\frac{s}{t}} \right)^{1 - q} }}{{\phi \left( q \right)}}.
\]
Here, all $a_i$ and $b_i$, $(i = 1,\cdots ,n)$ are assumed to be rational numbers.
Then, setting $c_i  = \frac{{l_i }}{{\sum\limits_{k = 1}^n {l_k } }},d_i  = \frac{{m_i }}{{\sum\limits_{k = 1}^n {m_k } }}$, $(i = 1,\cdots ,n)$  for some
natural numbers $l_i$ and $m_i$ such that $l_i \leq m_i$, and we set $
a_{i,j}  = \frac{1}{{\sum\limits_{k = 1}^n {l_k } }}\,\,\,\left(i = 1,\cdots ,n; {j = 1, \cdots ,l_i } \right),\,\,\,\,
b_{i,j}  = \frac{1}{{\sum\limits_{k = 1}^n {m_k } }}\,\,\left( i = 1,\cdots ,n; {j = 1, \cdots ,m_i } \right).$
Substituting into condition (R3), we have
\begin{eqnarray*}
&& \hspace*{-6mm}   D_q \left( \frac{1}{\sum\limits_{k = 1}^n {l_k }}, \cdots ,\frac{1}{\sum\limits_{k = 1}^n {l_k } },0, \cdots ,0, \cdots  \cdots ,\frac{1}{\sum\limits_{k = 1}^n {l_k } },  \right.\\
&&\left. \hspace*{6mm}   \cdots ,\frac{1}{\sum\limits_{k = 1}^n {l_k } },0, \cdots ,0 \left| \frac{1}{\sum\limits_{k = 1}^n {m_k } }, \cdots ,\frac{1}{\sum\limits_{k = 1}^n {m_k }} \right.  \right) \\
&& \hspace*{-6mm}    = D_q \left( {c_i , \cdots ,c_n \left| {d_1 , \cdots ,d_n } \right.} \right) \\
&& \hspace*{-6mm}   + \sum\limits_{i = 1}^n {c_i^q d_i^{1 - q} D_q \left( {\frac{1}{{l_i }}, \cdots ,\frac{1}{{l_i }},0, \cdots ,0\left| {\frac{1}{{ {m_i } }}, \cdots ,\frac{1}{{ {m_i } }}} \right.} \right)}. 
\end{eqnarray*}
Thus, we have
\begin{eqnarray*}
&& \hspace*{-6mm}   D_q \left( {c_i , \cdots ,c_n \left| {d_1 , \cdots ,d_n } \right.} \right) \\
&& = f_q \left( {\sum\limits_{k = 1}^n {l_k } ,\sum\limits_{k = 1}^n {m_k } } \right)  - \sum\limits_{i = 1}^n {c_i^q d_i^{1 - q} } f_q \left( {l_i ,m_i } \right) \\ 
&&  = \frac{{1 - \left( {\sum\limits_{k = 1}^n {l_k } } \right)^{1 - q} \left( {\sum\limits_{k = 1}^n {m_k } } \right)^{q - 1} }}{{\phi \left( q \right)}} \\
&&\hspace*{6mm}   - \frac{{\sum\limits_{i = 1}^n {c_i^q d_i^{1 - q} } \left( {1 - l_i ^{1 - q} m_i ^{q - 1} } \right)}}{{\phi \left( q \right)}} \\ 
&&  = \frac{{1 - \sum\limits_{i = 1}^n {c_i^q d_i^{1 - q} } }}{{\phi \left( q \right)}}.  
 \end{eqnarray*}
From condition (R1) and the fact that any real number can be approximated by a rational number, the above result holds for any positive real numbers $a_i$ and $b_i$
satisfying $\sum_{i=1}^na_i = \sum_{i=1}^n b_i =1$.

\hfill \qed
\vspace*{3mm}

We give a remark on the conditions of Axiom \ref{g_hobson_axiom} in the following proposition.

\begin{Prop}  \label{prop_relative01}
The following conditions (R3') and (R4), and the symmetry (R2) imply the condition (R3).
\begin{itemize}
\item[(R3')] {\it Generalized grouping axiom}: 
\begin{eqnarray*}
 &&\hspace*{-12mm}    D_q \left( a_{1,1} , \cdots ,a_{1,m} ,a_{2,1} , \cdots ,a_{2,m} \right.\\
&&\hspace*{-12mm}    \left. \hspace*{8mm}     | b_{1,1} , \cdots ,b_{1,m} ,b_{2,1} , \cdots ,b_{2,m}  \right) \\
&& \hspace*{-9mm}    = D_q \left( {c_1 ,c_2 \left| {d_1 ,d_2 } \right.} \right) \\ 
 && \hspace*{-6mm}    + c_1^q d_1^{1 - q} D_q \left( {\frac{{a_{1,1} }}{{c_1 }}, \cdots ,\frac{{a_{1,m} }}{{c_1 }}\left| {\frac{{b_{1,1} }}{{d_1 }}, \cdots ,\frac{{b_{1,m} }}{{d_1 }}} \right.} \right) \\
&&\hspace*{-6mm}     + c_2^q d_2^{1 - q} D_q \left( {\frac{{a_{2,1} }}{{c_2 }}, \cdots ,\frac{{a_{2,m} }}{{c_2 }}\left| {\frac{{b_{2,1} }}{{d_2 }}, \cdots ,\frac{{b_{2,m} }}{{d_2 }}} \right.} \right) 
 \end{eqnarray*}
where $c_i = \sum_{j=1}^m a_{i,j}$ and $d_i = \sum_{j=1}^m b_{i,j}$.
\item[(R4)] $D_q(A\vert B) =0$ if $a_j =b_j$ for all $j$.
\end{itemize}
\end{Prop}

{\it Proof:}
Here, the notation $\widetilde{D_q}(a_1,\cdots ,a_n \vert *) \equiv D_q(a_1,\cdots ,a_n \vert b_1 ,\cdots ,b_n)$ is introduced in order to simplify the proof.
From condition (R3'), we have
\begin{eqnarray}
&&  \hspace*{-12mm}   \widetilde{D_q }\left( {a_1 , \cdots ,a_n \left| * \right.} \right) = \widetilde{D_q }\left( {A,a_{n - 1}  + a_n \vert *} \right)  \nonumber  \\
&&    + A^q B^{1 - q} \widetilde{D_q }\left( {\frac{{a_1 }}{A}, \cdots ,\frac{{a_{n - 2} }}{A}\left| * \right.} \right)   \nonumber  \\ 
&&     + \left( {a_{n - 1}  + a_n } \right)^q \left( {b_{n - 1}  + b_n } \right)^{1 - q} \nonumber  \\ 
&& \hspace*{4mm}  \times  \widetilde{D_q }\left( {\frac{{a_{n - 1} }}{{a_{n - 1}  + a_n }},\frac{{a_n }}{{a_{n - 1}  + a_n }}\left| * \right.} \right)  
\label{pro_proof1}
\end{eqnarray}
and
\begin{eqnarray}
&& \hspace*{-14mm}   \widetilde{D_q }\left( {a_1 , \cdots ,a_{n - 2} ,a_{n - 1}  + a_n \left| * \right.} \right)  \nonumber   \\
&& \hspace*{-10mm}   = \widetilde{D_q }\left( {\left. {A,a_{n - 1}  + a_n } \right| *} \right)   \nonumber   \\ 
&&  \hspace*{-8mm}   + A^q B^{1 - q} \widetilde{D_q }\left( {\frac{{a_1 }}{A}, \cdots ,\frac{{a_{n - 2} }}{A}\left| * \right.} \right) \nonumber   \\
&&   \hspace*{-8mm}    + \left( {a_{n - 1}  + a_n } \right)^q \left( {b_{n - 1}  + b_n } \right)^{1 - q} D_q \left( {1\left| 1 \right.} \right) 
\label{pro_proof2}
\end{eqnarray}
where $A = a_1  +  \cdots  + a_{n - 2}$ and $B = b_1  +  \cdots  + b_{n - 2} $.
Thus the condition (R4) and the above equations Eq.(\ref{pro_proof1}) and Eq.(\ref{pro_proof2}) imply 
\begin{eqnarray}  
&& \hspace*{-18mm}   \widetilde{D_q }\left( {a_1 , \cdots ,a_n \left| * \right.} \right) \nonumber   \\ 
&& \hspace*{-12mm}   = \widetilde{D_q }\left( {a_1 , \cdots ,a_{n - 2} ,a_{n - 1}  + a_n \left| * \right.} \right)  \nonumber   \\ 
&& + \left( {a_{n - 1}  + a_n } \right)^q \left( {b_{n - 1}  + b_n } \right)^{1 - q} \nonumber   \\ 
&& \hspace*{4mm} \times  \widetilde{D_q }\left( {\frac{{a_{n - 1} }}{{a_{n - 1}  + a_n }},\frac{{a_n }}{{a_{n - 1}  + a_n }}\left| * \right.} \right)\, \label{pro_proof3}
\end{eqnarray}
Next, we derive condition (R3) by induction on $n$. Assuming that condition (R3) holds for some natural number $n$, 
we have
\begin{eqnarray}
&& \hspace*{-8mm}   \widetilde{D_q }\left( {a_{1,1} , \cdots ,a_{1,m} , \cdots ,a_{n + 1,1} , \cdots ,a_{n + 1,m} \left| * \right.} \right)  \nonumber     \\ 
&& \hspace*{-4mm}    = \widetilde{D_q }\left( {c_1 , \cdots ,c_{n - 1} ,c_n  + c_{n + 1} \left| * \right.} \right) \nonumber     \\ 
&& + \sum\limits_{i = 1}^{n - 1} {c_i^q d_i^{1 - q} \widetilde{D_q }\left( {\frac{{a_{i,1} }}{{c_i }}, \cdots ,\frac{{a_{i,m} }}{{c_i }}\left| * \right.} \right)} \nonumber    \\ 
&&  + \left( {c_n  + c_{n + 1} } \right)^q \left( {d_n  + d_{n + 1} } \right)^{1 - q}  \nonumber     \\ 
&&\hspace*{4mm} \times    \widetilde{D_q }\left( \frac{a_{n,1} }{c_n  + c_{n + 1} }, \cdots ,\frac{a_{n,m} }{c_n  + c_{n + 1} },   \right. \nonumber     \\ 
&&\left. \hspace*{12mm}      \frac{a_{n + 1,1} }{c_n  + c_{n + 1} }, \cdots ,\frac{a_{n + 1,m} }{c_n  + c_{n + 1} } | *  \right),  \label{pro_proof4}
\end{eqnarray}
by the use of (R2), (R4) and Eq.(\ref{pro_proof3}).
By Eq.(\ref{pro_proof3}), the first term in the right-hand side of Eq.(\ref{pro_proof4}) can be written as
\begin{eqnarray}
&& \hspace*{-14mm}   \widetilde{D_q }\left( {c_1 , \cdots ,c_{n - 1} ,c_n  + c_{n + 1} \left| * \right.} \right) \nonumber      \\
&& \hspace*{-8mm}    = \widetilde{D_q }\left( {c_1 , \cdots ,c_n ,c_{n + 1} \left| * \right.} \right)   \nonumber      \\ 
&&  - \left( {c_n  + c_{n + 1} } \right)^q \left( {d_n  + d_{n + 1} } \right)^{1 - q}\nonumber      \\
&& \hspace*{4mm} \times  \widetilde{D_q }\left( {\frac{{c_n }}{{c_n  + c_{n + 1} }},\frac{{c_{n + 1} }}{{c_n  + c_{n + 1} }}\left| * \right.} \right). \label{pro_proof5} 
\end{eqnarray}
By condition (R3'), the last term in the right-hand side of Eq.(\ref{pro_proof4}) can be written as
\begin{eqnarray}
&& \hspace*{-14mm}   \left( {c_n  + c_{n + 1} } \right)^q \left( {d_n  + d_{n + 1} } \right)^{1 - q} \nonumber       \\
&& \hspace*{-4mm}   \times\widetilde{D_q }\left( \frac{a_{n,1} }{c_n  + c_{n + 1} }, \cdots ,\frac{a_{n,m} }{c_n  + c_{n + 1} }, \right.  \nonumber       \\
&& \hspace*{4mm}   \left. \frac{a_{n + 1,1} }{c_n  + c_{n + 1} }, \cdots ,\frac{a_{n + 1,m} }{c_n  + c_{n + 1} } | *  \right)\nonumber       \\ 
&& \hspace*{-10mm} = \left( {c_n  + c_{n + 1} } \right)^q \left( {d_n  + d_{n + 1} } \right)^{1 - q}\nonumber       \\
&& \hspace*{-4mm} \times  \widetilde{D_q }\left( \frac{c_n }{c_n  + c_{n + 1} },\frac{c_{n + 1} }{c_n  + c_{n + 1} } | *  \right) \nonumber      \\ 
&&  + c_n^q d_n^{1 - q} \widetilde{D_q }\left( \frac{{a_{n,1} }}{{c_n }}, \cdots ,\frac{{a_{n,m} }}{{c_n }} | *  \right) \nonumber      \\ 
&& \hspace*{-8mm}      + c_{n + 1}^q d_{n + 1}^{1 - q} \widetilde{D_q }\left( \frac{{a_{n + 1,1} }}{{c_{n + 1} }}, \cdots ,\frac{{a_{n + 1,m} }}{{c_{n + 1} }} | * \right). \label{pro_proof6}
\end{eqnarray}
Substituting Eq.(\ref{pro_proof5}) and Eq.(\ref{pro_proof6}) into Eq.(\ref{pro_proof4}), we have  
\begin{eqnarray*}
&&\hspace*{-4mm} \widetilde{D_q }\left( {a_{1,1} , \cdots ,a_{1,m} , \cdots ,a_{n + 1,1} , \cdots ,a_{n + 1,m} \left| * \right.} \right)   \\ 
&& = \widetilde{D_q }\left( {c_1 , \cdots ,c_n ,c_{n + 1} \left| * \right.} \right) \\
&& \hspace*{4mm} + \sum\limits_{i = 1}^{n - 1} {c_i^q d_i^{1 - q} \widetilde{D_q }\left( {\frac{{a_{i,1} }}{{c_i }}, \cdots ,\frac{{a_{i,m} }}{{c_i }}\left| * \right.} \right)}  \\ 
&& \hspace*{4mm} + c_n^q d_n^{1 - q} \widetilde{D_q }\left( {\frac{{a_{n,1} }}{{c_n }}, \cdots ,\frac{{a_{n,m} }}{{c_n }}\left| * \right.} \right) \\
&& \hspace*{4mm} + c_{n + 1}^q d_{n + 1}^{1 - q} \widetilde{D_q }\left( {\frac{{a_{n + 1,1} }}{{c_{n + 1} }}, \cdots ,\frac{{a_{n + 1,m} }}{{c_{n + 1} }}\left| * \right.} \right) \\ 
\end{eqnarray*}
which means that condition (R3) holds for $n+1$. Thus, the proposition is proven.
\hfill \qed
\vspace*{3mm}

From Proposition \ref{prop_relative01}, we find that
we may adopt the axiom composed of the set of $\{$ (R1),(R2),(R3') and (R4) $\}$ instead of the set of $\{$ (R1),(R2) and (R3) $\}$ in Theorem \ref{the1}.

Condition (R3') of the present paper is a generalization of (H3) in the sense that the factor $c_i^qd_i^{q'}$ of the second term in the right-hand side of Eq.(\ref{Tsa_H3}) is 
a parametric extension of the factor $c_i^1d_i^0$ of that of Eq.(\ref{Hob_H3}), satisfying $q+q'=1$. Condition (R3') of the present study contains the information on $d_i$ as a factor of each
function $D_q$ in the second term in the right-hand side of Eq.(\ref{Tsa_H3}), whereas the original condition (H3) does not, because it is eliminated by $d_i^0 =1$ in such a special case.



\subsection{Characterization of $\phi(q)$}
In this subsection, 
the function $\phi(q)$ appeared in Theorem \ref{the1} is characterized.
\begin{Prop} \label{pro01}
The property whereby Tsallis relative entropy is one parametric extension of relative entropy:
\begin{equation} \label{cha1}
\lim_{q \to 1} D_q(A\vert B) = k \sum_{j=1}^n a_j \log \frac{a_j}{b_j}
\end{equation}
characterizes the function $\phi(q)$ such as
\begin{itemize}
\item[(c1)]  $\lim_{q \to 1}\phi(q) =0$.
\item[(c2)]  There exists an interval $(a,b)$ such that $a<1<b$, and $\phi(q)$ is differentiable on the interval $(a,1)\cup (1,b)$.
\item[(c3)]  There exists a positive number $k$ such that $\lim_{q \to 1} \frac{d\phi(q)}{dq} = -\frac{1}{k}$.
\end{itemize}
\end{Prop}

{\it Proof:}
Eq.(\ref{cha1}) implies condition (c1), since $\lim_{q \to 1} \left( 1- \sum_{i=1}^n a_i^qb_i^{1-q}  \right) =0$.
If we differentiate $1- \sum_{i=1}^n a_i^qb_i^{1-q} $ by $q$, then we obtain condition (c2).
Moreover, by l'Hopital's formula, we have 
$$  \lim_{q \to 1} D_q(A\vert B)=   \lim_{q\to 1} \frac{-\sum_{i=1}^n a_i^qb_i^{1-q}\left(\log a_i -\log b_i\right)}{\frac{d\phi(q)}{dq}} $$
which implies condition (c3).

\hfill \qed

\begin{Prop}\label{pro02}
The condition whereby $D_q(A\vert U)$ takes the minimum value for a fixed posterior probability distribution as a uniform distribution $U=\left\{\frac{1}{n},\cdots ,\frac{1}{n}  \right\}$ :
\begin{itemize}
\item[(R5)] {\it Minimization} :  $$D_q(a_1,\cdots ,a_n \vert \frac{1}{n},\cdots ,\frac{1}{n}) \geq D_q(\frac{1}{n},\cdots ,\frac{1}{n}\vert \frac{1}{n},\cdots ,\frac{1}{n})$$
\end{itemize}
implies 
\begin{itemize}
\item[(c4)] $\phi(q) (1-q) >0 $ for $q \neq 1$. 
\end{itemize}
\end{Prop}

{\it Proof:}
Since $1-n^{q-1} \sum_{i=1}^n a_i^q$ is concave in $a_i$ for $q >1$ and convex in $a_i$ for $0 \leq q <1$, we obtain condition (c4) in order to
satisfy the condition that $D_q(A\vert U)$ takes a minimum value.

\hfill \qed

As a simple example of $\phi(q)$ satisfying the conditions (c1) to (c4), we take $\phi(q) = 1-q$ and $k=1$, 
and then obtain the Tsallis relative entropy defined in Eq.(\ref{Tsallisrelative}). Also, setting $\phi(q) = 1-2^{1-q}$ and $k=1$, we obtain the relative entropy of type $\beta$
$$
D_q(A\vert B) = \frac{1-\sum_{j=1}^n a_j^q b_j^{1-q}}{1-2^{1-q}},
$$
which appeared in Eq. (7.2.46) of \cite{AD}. See \cite{RK,MK} for different approaches to the axiomatic characterization for the relative entropy of type $\beta$.


\section{Generalized Faddeev's axiom and uniqueness theorem for Tsallis entropy}   \label{sec5}
\subsection{Simplification of the uniqueness theorem for Tsallis entropy}
The function $S_q(x_1,\cdots ,x_n)$ is assumed to be defined for the $n$-tuple $(x_1, \cdots ,x_n)$ belonging to 
$\Delta _n $ and to take a value in $\hbox{R}^+ $.
In order to characterize the function $S_q(x_1,\cdots ,x_n)$, we introduce the following axiom, which is a slight generalization of the Faddeev's axiom.

\begin{Ax} {\bf (Generalized Faddeev's axiom)}\label{gfaddeev}
\begin{itemize}
\item[(F1)] {\it Continuity}: The function $f_q (x) \equiv S_q(x,1-x)$ with parameter $q\geq 0$ is continuous on the closed interval $\left[0,1\right]$
 and $f_q(x_0) > 0$ for some  $x_0 \in \left[0,1\right]$. 
\item[(F2)] {\it Symmetry}: 
For arbitrary permutation $\left\{x'_k\right\} \in \Delta_n$ of $\left\{x_k\right\} \in \Delta_n$,
\begin{equation}
S_q(x_1,\cdots ,x_n) = S_q(x'_1,\cdots ,x'_n).
\end{equation}
\item[(F3)] {\it Generalized additivity}:
For $x_n= y +z$, $y \geq 0$ and $z >0$,
\begin{eqnarray}
&& \hspace*{-22mm} S_q(x_1,\cdots ,x_{n-1},y,z) = S_q(x_1,\cdots ,x_n)  \nonumber \\
&& \hspace*{12mm} +x_n^q S_q \left(\frac{y}{x_n},\frac{z}{x_n}\right).
\end{eqnarray}
\end{itemize}
\end{Ax}

Conditions (F1) and (F2) are identical to the original Faddeev's conditions, except for the addition of the parameter $q$.
Condition (F3) is a generalization of the original Faddeev's additivity condition in the sense that condition (F3) of the present paper uses
$x_n^q$ as the factor of the second term in the right-hand side, whereas the original condition uses $x_n$ as this factor.
Note that condition (F3) of the present paper is a simplification of condition (G3) in \cite{Suy}, because condition 
(F3) of the present paper does not require the summation on $i$ from $1$ to $n$. Moreover, the present axiom does not require the maximality condition (G2) in \cite{Suy}.
Therefore, the present axiom improves the generalized Shannon-Khinchin's axiom introduced in \cite{Suy} for the characterization of Tsallis entropy.
For the above generalized Faddeev's axiom, we have the following uniqueness theorem for Tsallis entropy:

\begin{The}\label{the} 
Conditions (F1), (F2) and (F3) uniquely give the form of the function $S_q : \Delta_n \to \hbox{R}^+$ such that
\begin{equation} \label{the_eq0}
S_q(x_1,\cdots ,x_n) = - \lambda_q \sum_{i=1}^n x_i^q \ln_q x_i,
\end{equation}
where $\lambda_q$ is a positive constant that depends on the parameter $q \geq 0$.
\end{The}

{\it Proof:}
For the special case of $q=1$, the present theorem directly follows from the theorem shown in \cite{Tve}. Thus, $q \neq 1$ is assumed.
The proof of this theorem is similar to that presented by H. Tveberg \cite{Tve}.
From conditions (F2) and (F3), for any $x,y,z$ satisfying $x,y \geq 0, z >0$ and $x+y+z=1$, $S_q(x,y,z)$ is expanded into separate equations:
\begin{eqnarray*}
S_q(x,y,z) &=& S_q(x,y+z)+ (y+z)^q S_q \left(\frac{y}{y+z},\frac{z}{y+z}\right)  \\
&=& S_q(y,x+z) + (x+z)^q S_q\left(\frac{x}{x+z},\frac{z}{x+z}\right).
\end{eqnarray*} 
Therefore, we have
\begin{eqnarray}
&&\hspace*{-12mm}  f_q(x) + (1-x)^q f_q\left(\frac{y}{1-x}\right) \nonumber \\
&& = f_q(y) +(1-y)^q f_q\left(\frac{x}{1-y}\right) \label{the_eq1}
\end{eqnarray}
Since Eq.(\ref{the_eq1}) is defined for any $0\leq x <1$ and $0\leq y <1$, by setting $x=0$ and $y > 0$, we have
$$
f_q(0) +f_q(y) =f_q(y) +(1-y)^q f_q(0).
$$
Thus, we have 
\begin{equation} \label{ini_con1}
S_q(0,1)=f_q(0)=0.
\end{equation}
Integrating both sides of Eq.(\ref{the_eq1}) with respect to $y$ from $0$ to $1-x$, we have
\begin{eqnarray*}
&& \hspace*{-12mm} \int_0^{1 - x} {f_q \left( x \right)dy}  + \left( {1 - x} \right)^q \int_0^{1 - x} {f_q \left( {\frac{y}{{1 - x}}} \right)dy} \\
&&   \hspace*{-6mm}   = \int_0^{1 - x} {f_q \left( y \right)dy}  + \int_0^{1 - x} {\left( {1 - y} \right)^q f_q \left( {\frac{x}{{1 - y}}} \right)dy}, 
\end{eqnarray*} 
which can be deformed as follows:
\begin{eqnarray} 
&& \hspace*{-18mm} \left( {1 - x} \right)f_q \left( x \right) + \left( {1 - x} \right)^{q + 1} \int_0^1 {f_q \left( t \right)d} t \nonumber \\
&& \hspace*{-8mm}    = \int_0^{1 - x} {f_q \left( t \right)dt}  + x^{q + 1} \int_x^1 {t^{ - q - 2} f_q \left( t \right)dt}. \label{the_eq2}
\end{eqnarray}
Since the function $f_q(x)$ is continuous on the closed interval $\left[0,1\right]$ due to condition (F1), 
it is differentiable on the open interval $x \in (0,1)$. 
By differentiating both sides of Eq.(\ref{the_eq2}) and applying the relation
\begin{equation} \label{ini_con2}
f_q(x) = f_q(1-x)
\end{equation}
due to condition (F2), we have
\begin{eqnarray} 
&&\hspace*{-6mm}  \left( {1 - x} \right)f'_q \left( x \right) = \left( {q + 1} \right)\left( {1 - x} \right)^q \int_0^1 {f_q \left( t \right)d} t \nonumber \\
&& + \left( {q + 1} \right)x^q \int_x^1 {t^{ - q - 2} f_q \left( t \right)dt}  - \frac{{f_q \left( x \right)}}{x}.\label{the_eq3}
\end{eqnarray}
Since $f'_q(x)$ is also differentiable on $(0,1)$, by again differentiating both sides of Eq.(\ref{the_eq3}), we have
\begin{eqnarray}
&&\hspace*{-6mm}  \left( {1 - x} \right)f''_q \left( x \right) =  - q\left( {q + 1} \right)\left( {1 - x} \right)^{q - 1} \int_0^1 {f_q \left( t \right)d} t \nonumber \\
&& + q\left( {q + 1} \right)x^{q - 1} \int_x^1 {t^{ - q - 2} f_q \left( t \right)dt}  \nonumber \\ 
 && - \frac{{qf_q \left( x \right)}}{{x^2 }} - \frac{{f'_q \left( x \right)}}{x} + f'_q \left( x \right) \label{the_eq4} 
\end{eqnarray}
Multiplying both sides of Eq.(\ref{the_eq4}) by $x$, we have
\begin{eqnarray}
&& \hspace*{-8mm} x\left( {1 - x} \right)f''_q \left( x \right)    =  - q\left( {q + 1} \right)x\left( {1 - x} \right)^{q - 1} \int_0^1 {f_q \left( t \right)d} t  \nonumber \\ 
&& \hspace*{4mm} + q\left( {q + 1} \right)x^q \int_x^1 {t^{ - q - 2} f_q \left( t \right)dt}  \nonumber \\ 
 && \hspace*{4mm} - \frac{{qf_q \left( x \right)}}{x} + \left( {x - 1} \right)f'_q \left( x \right). \label{the_eq5}
\end{eqnarray}
Also, multiplying both sides of Eq.(\ref{the_eq3}) by $q$, we obtain 
\begin{eqnarray} 
&& \hspace*{-4mm} q\left( {1 - x} \right)f'_q \left( x \right) = q\left( {q + 1} \right)\left( {1 - x} \right)^q \int_0^1 {f_q \left( t \right)d} t \nonumber \\
&& + q\left( {q + 1} \right)x^q \int_x^1 {t^{ - q - 2} f_q \left( t \right)dt}  - \frac{{qf_q \left( x \right)}}{x}. \label{the_eq6}
\end{eqnarray}
From Eq.(\ref{the_eq5}) and Eq.(\ref{the_eq6}), we have the following differential equation:
\begin{equation} \label{the_eq7}
xf''_q \left( x \right) = \left( {q - 1} \right)f'_q \left( x \right) - q \lambda _q \left( {1 - x} \right)^{q - 2}, 
\end{equation}
where $\lambda_q \equiv (q+1) \int_0^1 {f_q \left( t \right)d} t$.
This differential equation has the following general solution with constants $c_1(q)$ and $c_2(q)$, which depend only on parameter $q$:
\begin{equation} \label{the_eq8}
f_q \left( x \right) = c_1(q)  + c_2(q) x^q + \frac{{\lambda _q \left( {1 - x} \right)^q }}{{1 - q}}.
\end{equation}
The initial condition Eq.(\ref{ini_con1}) implies $c_1(q) = \frac{\lambda_q}{q-1}$. In addition, the condition Eq.(\ref{ini_con2}) implies
$c_2(q) = \frac{\lambda_q}{1-q} $.
Substituting $c_1(q)$ and $c_2(q)$ into Eq.(\ref{the_eq8}), we have
\begin{equation} \label{the_eq9}
 f_q(x) = -\lambda_q \left\{ x^q \ln_q x +(1-x)^q \ln_q (1-x) \right\},
\end{equation}
after the calculations.
Since there exists some $t_0 \in [0,1]$ such that $f_q(t_0) > 0$, due to condition (F1), and the range of $S_q$ is $\hbox{R}^+$, we have $\lambda_q >0$ for any $q \geq 0$.
Therefore, Eq.(\ref{the_eq0}) is proven for $n=2$. Finally, Eq.(\ref{the_eq0}) is proven for the general $n \geq 3$ by induction on $n$.
Assuming that Eq.(\ref{the_eq0}) is true for any $n \in \hbox{N}$, we obtain the following calculations:
\begin{eqnarray*}
 && \hspace*{-4mm}  S_q \left( {x_1 , \cdots ,x_n ,x_{n + 1} } \right) = S_q \left( {x_1 , \cdots ,x_n  + x_{n + 1} } \right) \\
&& \hspace*{2mm}  + \left( {x_n  + x_{n + 1} } \right)^q S_q \left( {\frac{{x_n }}{{x_n  + x_{n + 1} }},\frac{{x_{n + 1} }}{{x_n  + x_{n + 1} }}} \right) \\ 
 && =  - \lambda _q \sum\limits_{i = 1}^{n - 1} {x_i^q \ln _q x_i }  - \lambda _q \left( {x_n  + x_{n + 1} } \right)^q \ln _q \left( {x_n  + x_{n + 1} } \right) \\ 
 && \hspace*{1mm}  - \lambda _q \left( {x_n  + x_{n + 1} } \right)^q \left\{ \left( \frac{x_n }{x_n  + x_{n + 1} } \right)^q \ln _q \left( \frac{x_n }{x_n  + x_{n + 1} } \right) \right. \\
&&\left.\hspace*{8mm}   + \left( \frac{x_{n + 1} }{x_n  + x_{n + 1} } \right)^q \ln _q \left( \frac{x_{n + 1} }{x_n  + x_{n + 1} } \right) \right\} \\ 
 && =  - \lambda _q \sum\limits_{i = 1}^{n - 1} {x_i^q \ln _q x_i }    - \lambda _q \left( {x_n  + x_{n + 1} } \right)^q \ln _q \left( {x_n  + x_{n + 1} } \right) \\ 
 && \hspace*{1mm}  - \lambda _q x_n^q \ln _q \left( {\frac{{x_n }}{{x_n  + x_{n + 1} }}} \right)   - \lambda _q x_{n + 1}^q \ln _q \left( {\frac{{x_{n + 1} }}{{x_n  + x_{n + 1} }}} \right) \\ 
 && =  - \lambda _q \sum\limits_{i = 1}^{n - 1} {x_i^q \ln _q x_i }  + \lambda _q \left( {x_n  + x_{n + 1} } \right)\ln _q \frac{1}{{x_n  + x_{n + 1} }} \\ 
 && \hspace*{4mm}  - \lambda _q x_n^q \left( {\ln _q x_n  + x_n^{1 - q} \ln _q \frac{1}{{x_n  + x_{n + 1} }}} \right)\\
&& \hspace*{4mm}   - \lambda _q x_{n + 1}^q \left( {\ln _q x_{n + 1}  + x_{n + 1}^{1 - q} \ln _q \frac{1}{{x_n  + x_{n + 1} }}} \right) \\ 
 && =  - \lambda _q \sum\limits_{i = 1}^{n + 1} {x_i^q \ln _q x_i }.   
\end{eqnarray*}
This shows that Eq.(\ref{the_eq0}) is also true for $n+1$. Thus, the proof of this theorem is completed.

\hfill \qed

\begin{Rem}
If further conditions are imposed on Axiom \ref{gfaddeev} such that $S_q(\frac{1}{2},\frac{1}{2}) =1$, which is the normalization condition to characterize
the structural $a$-entropy in \cite{HC}, then we obtain $\lambda_q = \frac{1}{\ln_q 2}$ from Eq.(\ref{the_eq9}).
Thus, we straightforwardly obtain the structural $a$-entropy \cite{HC}:
$$
S_q(x_1, \cdots , x_n)=\frac{1-\sum_{i=1}^n x_i^q}{1-2^{1-q}},
$$
in a manner similar to the induction of the proof in Theorem \ref{the}. 
This means that the present theorem includes the Havrda-Charv\'at's axiom as a special case. 
Note that the present axiom requires the symmetry condition, whereas the Havrda-Charv\'at's axiom requires the expandability condition.
\end{Rem}


\subsection{A relation to the generalized Shannon-Khinchin's axiom}
Finally, the relationship between the generalized Shannon-Khinchin's axiom introduced in \cite{Suy} and the generalized Faddeev's axiom is studied.

\begin{Prop}\label{pro1}
Axiom \ref{gen_SK} implies Axiom \ref{gfaddeev}.
\end{Prop}

{\it Proof:}
The fact that conditions (G1) and (G2) imply condition (F1) is trivial, thus we show conditions (G1) and (G3) imply condition (F2).
If all $x_i$, $(i=1,\cdots ,n)$ are positive rational numbers, then each $x_i$ can be represented by $\frac{l_i}{m}$, $(2 \leq l_i \leq m, \,\, l_i,m \in \hbox{Z})$.
Applying condition (G3), since $x_i=\frac{l_i}{m} = \sum_{j=1}^{l_i} \frac{1}{m}$, we have
\begin{eqnarray*}
&& \hspace*{-12mm}  S_q \left( {x_1 , \cdots ,x_n } \right) = S_q \left( {\frac{{l_1 }}{m}, \cdots ,\frac{{l_n }}{m}} \right) \\ 
&&  = S_q \left( {\underbrace {\frac{1}{m}, \cdots ,\frac{1}{m}}_{l_1 }, \cdots ,\underbrace {\frac{1}{m}, \cdots ,\frac{1}{m}}_{l_n }} \right) \\
&& \hspace*{4mm}  - \sum\limits_{i = 1}^m {x_i^q S_q \left( {\frac{1}{{l_i }}, \cdots ,\frac{1}{{l_i }}} \right)}. 
\end{eqnarray*}
The first term in the right-hand side of the above equation does not depend on the order of $(l_1,\cdots ,l_n)$. 
In addition, the method how to take the summation in the second term of the right-hand side
of the above equation is arbitrary, so that 
the above equation is equivalent to
$$  S_q \left( {\underbrace {\frac{1}{m}, \cdots ,\frac{1}{m}}_{l'_1 }, \cdots ,\underbrace {\frac{1}{m}, \cdots ,\frac{1}{m}}_{l'_n }} \right) 
   - \sum\limits_{i = 1}^m {x'_i\,\! ^q S_q \left( {\frac{1}{{l'_i }}, \cdots ,\frac{1}{{l'_i }}} \right)} $$
for the permutation $\left\{x'_i\right\}$ from $\left\{x_i \right\}$ where $x'_i = \frac{l'_i}{m}$, $(2\leq l'_i \leq m, \,\, l'_i,m\in\hbox{Z})$.
That is, condition (F2) holds for any rational numbers $x_i$.
If $x_i$ is not a rational number, then the continuity of condition (G1) after the approximation of $x_i$ by a rational number is used, and then we have condition (F2).
Finally, conditions (G3) and (G4) are applied to imply (F3).
From conditions (G3), (G4) and (F2), we have
\begin{eqnarray*}
&& \hspace*{-4mm}  S_q \left( {\frac{1}{2},\frac{1}{2}} \right) = S_q \left( {\frac{1}{2},\frac{1}{2},0,0} \right) = S_q \left( {\frac{1}{2},0,\frac{1}{2},0} \right) \\ 
&&  = S_q \left( {\frac{1}{2},\frac{1}{2}} \right) + \frac{1}{{2^q }}S_q \left( {1,0} \right) + \frac{1}{{2^q }}S_q \left( {1,0} \right) . 
\end{eqnarray*}
We therefore obtain $S_q(1,0)=0$, and thus it follows that 
\begin{eqnarray*}
&&\hspace*{-8mm}   S_q \left( {x_1 , \cdots ,x_{n - 1} ,y,z} \right) \\
&& = S_q \left( {x_1 ,0,x_2 ,0, \cdots ,x_{n - 1} ,0,y,z} \right) \\ 
&&  = S_q \left( {x_1 , \cdots ,x_n } \right) + \sum\limits_{i = 1}^{n - 1} {x_i^q S_q \left( {1,0} \right)}  + x_n^q S_q \left( {\frac{y}{{x_n }},\frac{z}{{x_n }}} \right) \\ 
&&  = S_q \left( {x_1 , \cdots ,x_n } \right) + x_n^q S_q \left( {\frac{y}{{x_n }},\frac{z}{{x_n }}} \right),  
\end{eqnarray*}
which implies condition (F3).

\hfill \qed

In addition, we have the following proposition:

\begin{Prop}\label{pro2}
$S_q(X) =-\lambda_q \sum_{i=1}^n x_i^q \ln_q x_i$ satisfies Axiom \ref{gen_SK}.
\end{Prop}

{\it Proof:}
Conditions (G1) and (G4) are trivial. 
Condition (G2) is proven using the non-negativity of the Tsallis relative entropy:
$$
D_q(X\vert Y) \equiv - \sum_{i=1}^n  x_i \ln_q \frac{y_i}{x_i}
$$ 
for two random variables $X$ and $Y$, where $\left\{x_i\right\}$ and $\left\{y_i\right\}$, $(i=1,\cdots ,n)$ are probability distributions of $X$ and $Y$, respectively.
See \cite{FYK} for the mathematical properties of the Tsallis relative entropy.
Its non-negativity can be easily proven by the convexity of the function $-\ln_q(x)$. The non-negativity implies $S_q(X) \leq \ln_q n$
by setting a random variable $U=\left\{\frac{1}{n},\cdots ,\frac{1}{n} \right\}$ having a uniform distribution, as a special case of $Y$.
The maximum value is attained when $X = \left\{\frac{1}{n}, \cdots \frac{1}{n} \right\}$.
Note that $\lambda_q$ does not depend on the method used to take the maximum of $S_q(X)$. Thus, condition (G2) is proven.
Finally, condition (G3) is proven by direct calculation.
\begin{eqnarray*}
&& \hspace*{-6mm}  S_q \left( {x_1 , \cdots ,x_n } \right) + \sum\limits_{i = 1}^n {x_i^q S_q \left( {\frac{{x_{i1} }}{{x_i }}, \cdots ,\frac{{x_{im_i } }}{{x_i }}} \right)}  \\ 
&&  \hspace*{-4mm}  =  - \lambda _q \sum\limits_{i = 1}^n {x_i^q \ln _q x_i }  + x_1^q S_q \left( {\frac{{x_{11} }}{{x_1 }}, \cdots ,\frac{{x_{1m_1 } }}{{x_1 }}} \right) \\
&& \hspace*{-4mm}  +  \cdots  + x_n^q S_q \left( {\frac{{x_{n1} }}{{x_n }}, \cdots ,\frac{{x_{nm_n } }}{{x_n }}} \right) \\ 
&&  \hspace*{-4mm}  =  - \lambda _q \sum\limits_{i = 1}^n {x_i^q \ln _q x_i }  \\
&& \hspace*{-4mm}  - \lambda _q x_1^q \left\{ \left( {\frac{x_{11} }{x_1 }} \right)^q \ln _q \frac{{x_{11} }}{{x_1 }} +  \cdots  + \left( {\frac{{x_{1m_1 } }}{{x_1 }}} \right)^q \ln _q \frac{x_{1m_1 } }{x_1 } \right\} \\ 
&& \hspace*{-4mm}  -  \cdots   \\ 
&&  \hspace*{-4mm}  - \lambda _q x_n^q \left\{ \left( {\frac{{x_{n1} }}{{x_n }}} \right)^q \ln _q \frac{{x_{n1} }}{{x_n }} +  \cdots + \left( {\frac{{x_{nm_n } }}{{x_n }}} \right)^q \ln _q \frac{{x_{nm_n } }}{{x_n }} \right\} \\ 
&& \hspace*{-4mm}   =  - \lambda _q \sum\limits_{i = 1}^n {x_i^q \ln _q x_i } 
- \lambda _q \left( x_{11}^q \ln _q x_{11}  + x_{11} \ln _q \frac{1}{{x_1 }} +  \cdots \right.\\
&& \left. \hspace*{12mm}  + x_{1m_1 }^q \ln _q x_{1m_1 }  + x_{1m_1 } \ln _q \frac{1}{{x_1 }} \right) -  \cdots  \\ 
&&  \hspace*{4mm}  - \lambda _q \left( x_{n1}^q \ln _q x_{n1}  + x_{n1} \ln _q \frac{1}{{x_n }} +  \cdots \right. \\
&& \left. \hspace*{12mm}  + x_{nm_1 }^q \ln _q x_{nm_1 }  + x_{nm_1 } \ln _q \frac{1}{{x_n }} \right) \\ 
&&  \hspace*{-4mm}  = \lambda _q \left( {x_1 \ln _q \frac{1}{{x_1 }} +  \cdots  + x_n \ln _q \frac{1}{{x_n }}} \right) \\
&& \hspace*{4mm}  + S_q \left( {x_{11} , \cdots ,x_{1m_1 } , \cdots ,x_{n1} , \cdots ,x_{nm_n } } \right) \\ 
&&  \hspace*{4mm}  - \lambda _q \left\{ \left( {x_{11}  +  \cdots  + x_{1m_1 } } \right)\ln _q \frac{1}{{x_1 }} +  \cdots \right. \\
&&\left. \hspace*{12mm}  + \left( {x_{n1}  +  \cdots  + x_{nm_n } } \right)\ln _q \frac{1}{{x_n }} \right\} \\ 
&&  \hspace*{-4mm}  = S_q \left( {x_{11} , \cdots ,x_{1m_1 } , \cdots ,x_{n1} , \cdots ,x_{nm_n } } \right). 
\end{eqnarray*}

\hfill \qed

From Theorem \ref{the}, Proposition \ref{pro1} and  Proposition \ref{pro2}, we have the following equivalent relation among Axiom \ref{gfaddeev}, Axiom \ref{gen_SK} and Tsallis entropy:

\begin{The} 
The following three statements are equivalent.
\begin{itemize}
\item[(1)] $S_q : \Delta_n \to \hbox{R}^+$ satisfies Axiom \ref{gen_SK}.
\item[(2)] $S_q : \Delta_n \to \hbox{R}^+$ satisfies Axiom \ref{gfaddeev}.
\item[(3)] For $(x_1,\cdots ,x_n) \in \Delta_n$, there exists $\lambda_q >0$ such that  
$$S_q(x_1,\cdots ,x_n) =-\lambda_q \sum_{i=1}^n x_i^q \ln_q x_i. $$
\end{itemize}
\end{The}


\section{Conclusions}  \label{sec6}
The uniqueness theorem for Tsallis entropy introduced by Suyari \cite{Suy} was generalized to the case of Tsallis relative entropy and was simplified according to the manner of Faddeev \cite{Fad,Tve}.

Tsallis relative entropy was characterized by the generalized Hobson's axiom. The present result includes the uniqueness theorem proven by Suyari as a special case, 
in the sense that the choice of a trivial distribution for $B=\left\{b_j \right\}$ of Tsallis relative entropy produces 
the essential form of Tsallis entropy.
However, note that the present theorem requires the symmetry of condition (R2), whereas that of Suyari does not.

Moreover, Tsallis entropy was characterized by the generalized Faddeev's axiom, 
which is a simplification of the generalized Shannon-Khinchin's axiom introduced in
\cite{Suy}, and the uniqueness theorem proved in \cite{Suy} was slightly improved by introducing the generalized Faddeev's axiom. 
At the same time, the present result provides a generalization of the uniqueness theorem for Shannon entropy by means of the Faddeev's axiom.

\section*{Acknowledgement}
The author would like to thank Professor H. Suyari for providing the opportunity to read his paper \cite{Suy} before publication.
The author would also like to thank the reviewers for providing valuable comments.

\begin{biography}{Shigeru FURUICHI}
received the B.S. degree from Department of Mathematics, Tokyo University of Science, and M.S. and Ph.D. from Department of information Science,
  Tokyo University of Science, Japan. He was an Assistant Professor during 1997-2001 and has been a Lecturer from 2001 
in Department of Electronics and Computer Science,
Tokyo University of Science, Yamaguchi, Japan. His research interests include information theory, entropy theory and operator theory.
\end{biography}
\end{document}